# Incidence Handling and Response System

(make the computer network more secure and capable enough to withstand any kind of attack.)

Prof.Dhananjay R.Kalbande
Asst.Professor, Dept.of Computer Engg..
Sardar Patel Institute Of Technology (SPIT),
Mumbai,India.
E-mail:k_dhananjay@yahoo.com

Mr.Manish Singh
Student
Sardar Patel Institute of Technology(SPIT),
Mumbai,India.
E-mail:manishspit@yahoo.co.in

Prof.Dr.G.T.Thampi
Principal
Pillai's Institute of Information Technology (PIIT),
New Panvel,India
E-mail:gtthampi@yahoo.com

*Abstract*-**A computer network can be attacked in a number of ways. The security-related threats have become not only numerous but also diverse and they may also come in the form of blended attacks. it becomes difficult for any security system to block all types of attacks. this gives rise to the need of an incidence handling capability which is necessary for rapidly detecting incidents, minimizing loss and destruction, mitigating the weaknesses that were exploited and restoring the computing services. incidence response has always been an important aspect of information security but it is often overlooked by security administrators. in this paper, we propose an automated system which will handle the security threats and make the computer network capable enough to withstand any kind of attack. we also present the state-of-the-art technology in computer, network and software which is required to build such a system.**

I. INTRODUCTION

In an organization, thousands of possible signs of incidents may occur each day. The organizations that are attacked typically call a Computer Security and Incident Response Team(CSIRT) to handle the incident.However, CSIRTs are not accessible to all . A system can be built which utilizes the services of host-based and network-based IDPSs, firewalls,antivirus and other tools to detect incidents, auditing and forensic tools to gather evidence of incident,implements event correlation and centralized logging to intelligently analyze and classify incidents,mitigate incidents and restore the system to a stable state.We aim at developing such an automated system.

We make use of the required softwares and tools to build the system in the following way:

- We convert Snort which is an network IDS into an adaptive IDS/IPS .
- We make use of host-based IDS i.e Tripwire to ensure the integrity of critical system files and directories by identifiying all changes made to them.
- We use the sleuth kit to examine the file system and layout of disks and other media of the suspected computer
- We install Clam AntiVirus toolkit to scan the files and emails too using command line.
- We make use of SSH protocol to establish a secure connection between two systems
- We use Tcpdump to sniff packets over network.

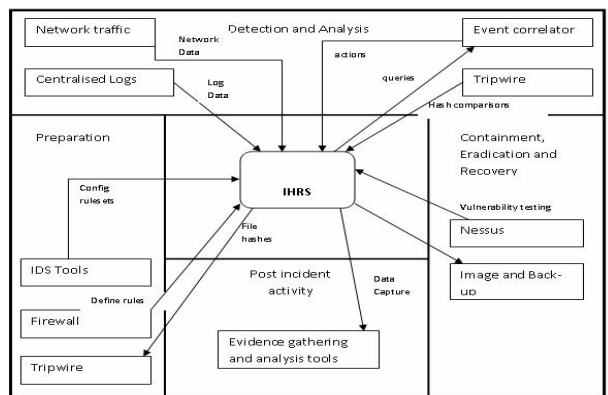

Figure 1.System Context Diagram Of Incidence Handling And Response System

## II. SOFTWARES AND TOOLS TO BE USED

### A. Network Intrusion Detection System (NIDS)

NIDS are intrusion detection systems that capture data packets traveling on the network media (cables, wireless) and match them to a database of signatures.

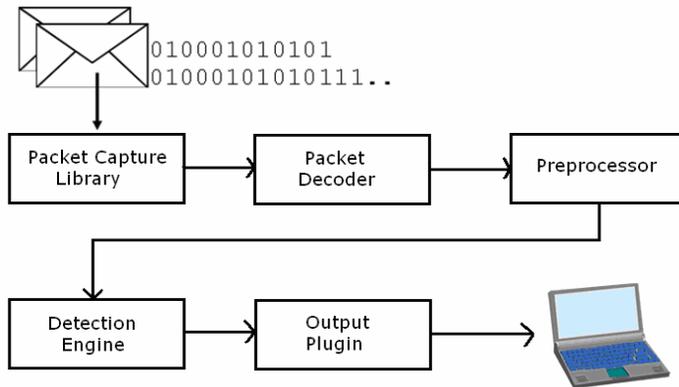

Figure 2.Architecture Of Network Intrusion Detection System

Depending upon whether a packet is matched with an intruder signature, an alert is generated or the packet is logged to a file or database.

**Packet Capture Library:**
Supports raw packet capture (packet sniffing).
**Packet Decoder:**
Once captured, packets are passed to a packet decoder, so that they can be translated into an IDS's internal data structure that provide a uniform basis for packet analysis.
**Preprocessor:**
Preprocessors are very important for any IDS to prepare data packets to be analyzed against rules in the detection engine.
**The Detection Engine:**
The detection engine is the most important part of an IDS. Its responsibility is to detect if any intrusion activity exists in a packet.
**Output Plug-in:**
If suspicious activity is identified by the Detection Engine, output plug-ins are called to generate administrative alerts.

Now in this project , we configure snort and run it in IDS mode so that the packets or alerts are logged into a file say alerts.ids in /var/log/snort directory.
We make use of classification.config located in /etc/snort to filter the log file alerts.ids and separate the ipaddresses from where the attacks such as portscan or ICMP flooding attack were launched.We perform the above operation by writing a shell script which will block the attacker's ipaddress using the command:
Route –add host ipaddress reject;
The ICMP ping flooding can be stopped by using the following command for iptables:
Iptables –A input

Thus snort detects all possible types of attacks and classifies them using the classification.config file. The shell script matches the alerts with the classification , classifying the attacks and acts accordingly. Thus in this way we build an adaptive IDS/IPS system that will block all major types of attacks like Denial of Service, Trojan, portscans and others.

### B. Host Intrusion Detection System (HIDS)

Host-based intrusion detection systems or HIDS are installed as agents on a host. These intrusion detection systems can look into system and application log files to detect any intruder activity. Some of these systems are reactive, meaning that they inform you only when something has happened. Some HIDS are proactive; they can sniff the network traffic coming to a particular host on which the HIDS is installed and alert you in real time.

*Open Source Tripwire*

Tripwire is a program that helps Linux Administrators detect "unauthorized activity" on their computers by creating a baseline database (digitally signed by you using a Passphrase that you mention during installation of the product) of all the files on the computer, including information about each file's size and last modification date. It generates the baseline by taking a snapshot of specified files and directories in a known secure state. (For maximum security, Tripwire should be installed and the baseline created before the system is at risk from intrusion.) Once it is generated, you can then check the current state of your system against the original baseline database and get a report of all the files that have been modified, deleted or added. The baseline database consists of checksum numbers generated by encryption algorithms from the file's contents.
Using Tripwire for intrusion detection and damage assessment helps you keep track of system changes and can speed the recovery from a break-in by reducing the number of files you must restore to repair the system.

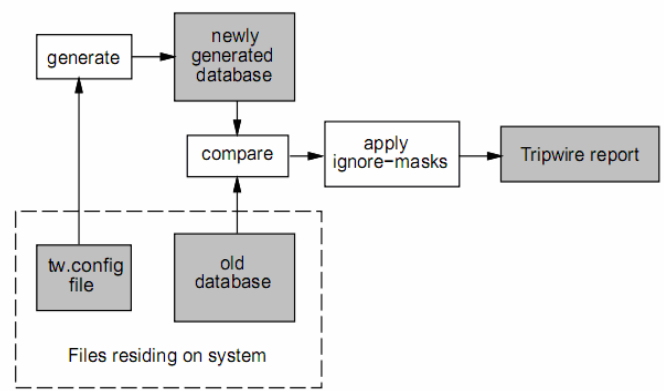

Figure 3. Working of tripwire

*Modes of operation*

Tripwire runs in one of four modes: Database Generation, Database Update, Integrity Checking, or Interactive Update mode.
*Database Generation*
$ tripwire --init
The /usr/local/lib/tripwire directory contains the Tripwire database of your system's files (*.twd) named as host.twd, where host is the host-name of the linux machine and a report directory where Tripwire reports are stored.
*Database Update*
$ tripwire -–update
*Integrity check*
$ tripwire -–check > /tmp/report.txt
It will then proceed to check your filesystem against the database and will create a file called report.txt in /tmp which will contain information on what Tripwire discovered.

*C. The Sleuth Kit*

The Sleuth Kit (previously known as TASK) is a collection of UNIX-based command line file and volume system forensic analysis tools. The file system tools allow you to examine file systems of a suspect computer in a non-intrusive fashion. Because the tools do not rely on the operating system to process the file systems, deleted and hidden content is shown. The volume system (media management) tools allow you to examine the layout of disks and other media. The Sleuth Kit supports DOS partitions, BSD partitions (disk labels), Mac partitions, Sun slices (Volume Table of Contents), and GPT disks. With these tools, you can identify where partitions are located and extract them so that they can be analyzed with file system analysis tools

Timelines:
Creating a timeline of system activity will give an investigator clues regarding where to probe further. The timelines in The Sleuth Kit allow one to quickly get a high-level look at system activity, such as when files were compiled and when archives were opened. TSK allows you to generate timelines of activity from a variety of sources. Autopsy allows you to also create timelines using the TSK tools. At a high level, timeline generation is a two step process. In the first step, temporal data is gathered from various data sources (such as file systems, registries, logs, etc.) and saved to the body file format. This step is done using the 'fls' tool in TSK or other tools, which are listed below. The second step is to sort and merge all of the temporal data into a single timeline. This step is done using the 'mactime' script in TSK.

- Data Gathering
  The primary method for collecting temporal data from file systems is to run fls with the '-m' flag. The fls command saves its output to a file in the body file format.
  Sample fls command:

  # fls -m "C:/" -o 63 -r images/disk.dd > body.txt

- Mactime
  mactime creates an ASCII timeline of file activity based on the output of the fls tool. It can be used to detect anomalous behavior and reconstruct events. mactime reads the body file (using the '-b' argument), which contains a line for each file or event. mactime then sorts the data based on its temporal data and prints the result. It can optionally use a starting date or a date range to limit the data being printed.
  Sample mactime output:

  # mactime -b body.txt 2002-03-01 > tl.03.01.2002.txt
  body.txt
  Date/Time Size  Activity Unix  User Group inode File Name
          (Bytes) Type   Permissions  Id Id
  Example:
  [...]
  Thu Aug 21 2003 01:20:38 512 m.c. -/-rwxrwxrwx 0 0 4 /file1.dat 900 m.c. -/-rwxrwxrwx 0 0 8 /file3.dat
  Thu Aug 21 2003 01:21:36 512 m.c. -/-rwxrwxrwx 0 0 12 /_ILE5.DAT (deleted)
  21 Thu Aug 2003 01:22:56 512 .a.. -/-rwxrwxrwx 0 0 4 /file1.dat
  [...]

*D. The Clam Antivirus Toolkit*

Clam AntiVirus is an open source (GPL) anti-virus toolkit for UNIX, designed especially for e-mail scanning on mail gateways. It provides a number of utilities including a flexible and scalable multi-threaded daemon, a command line scanner and advanced tool for automatic database updates. The important facilities provided by ClamAV, relevant to this project are:

- Licensed under the GNU General Public License, Version 2

- Command-line scanner

- Fast, multi-threaded daemon with support for on-access scanning

- Scans within archives and compressed files (also protects against archive bombs), including zip, rar, arj, gzip, bzip2, etc.

- Support for other special file formats including rtf, doc, pdf, etc.

- Advanced database updater with support for scripted updates and digital signatures

- Virus database updated multiple times per day

For example, to scan all files and sub-directories in the folder '/home', use the following command:

  clamscan -r /home

Some return codes of the command, relevant to the project, are:

0 : No virus found.

1 : Virus(es) found.

54: Can't open file.

*Freshclam:*

**Freshclam**.is the Clam AntiVirus Database Updater. It is configured by the file freshclam.conf. This file has to be modified by the user in the appropriate way for freshclam to work properly. Some lines that need to be modified in freshclam.conf look like this after modification:

DatabaseDirectory /var/lib/clamav
UpdateLogFile /var/log/clamav/freshclam.log
DatabaseOwner clamav
DNSDatabaseInfo current.cvd.clamav.net
DatabaseMirror db.in.clamav.net
DatabaseMirror database.clamav.net
MaxAttempts 3
Checks 24
NotifyClamd /etc/clamav.conf

*E. SSH protocol*

SSH allows users to log into host systems remotely. Unlike rlogin or telnet SSH encrypts the login session, making it impossible for intruders to collect clear-text passwords. SSH is designed to replace common methods for remotely logging into another system through a command shell. A related program called scp replaces older programs designed to copy files between hosts such as ftp or rcp. Using secure methods to remotely log in to other systems will decrease the security risks for both your system and the remote system. SSH (or *S*ecure *SH*ell) is a protocol for creating a secure connection between two systems. In the SSH protocol, the client machine initiates a connection with a server machine. The following safeguards are provided by SSH:

After an initial connection, the client verifies it is connecting to the same server during subsequent sessions.

- The client transmits its authentication information to the server, such as a username and password, in an encrypted format.

- All data sent and received during the connection is transferred using strong, 128 bit encryption, making it extremely difficult to decrypt and read.

- The client has the ability to use X11 applications launched from the shell prompt. This technique, called *X11 forwarding*, provides a secure means to use graphical applications over a network. Because the SSH protocol encrypts everything it sends and receives, it can be used to secure otherwise insecure protocols. Using a technique called *port forwarding*, an SSH server can become a conduit to secure insecure protocols, like POP, increasing overall system and data security.

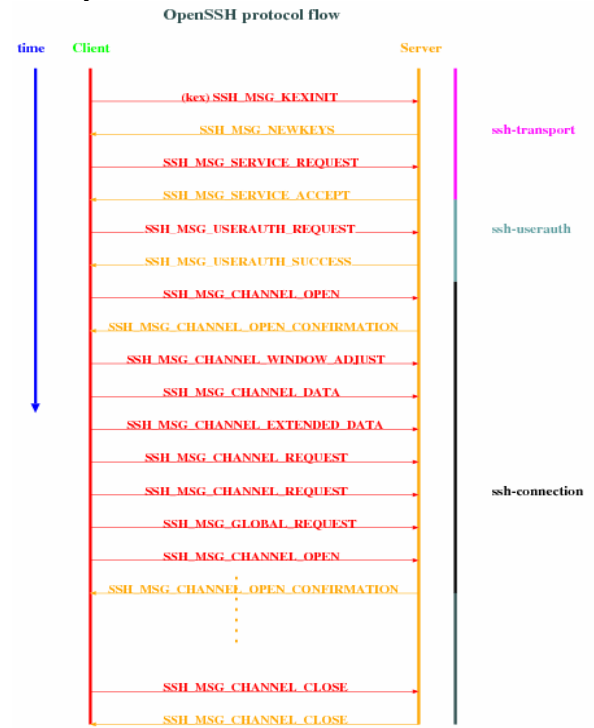

Figure 4. The SSH protocol

## III. IMPLEMENTATION

To reduce the load on the machine when the user might be doing some other work, and to facilitate a semi-automation of the system, the scripts to run snort, tripwire, antivirus, chkrootkit, and backups are added to crontabs. In Linux, crontabs are located at /var/spool/cron/tabs. Whatever scripts are added to this file are executed periodically, as specified .

There are five fields representing the time / periodicity of execution. The fields represent different date parts in the following order:

1. minute (from 0 to 59)

2. hour (from 0 to 23)

3. day of month (from 1 to 31)

4. month (from 1 to 12)

5. day of week (from 0 to 6) (0=Sunday)

A "*" in any of the field signifies "for every value". The following script semi-automates the system. It adds jobs to crontabs. It should be executed with root privilege.

```
#! /bin/sh
# automate.sh

echo "snort –dev –c /etc/snort/snort.conf –l /var/log/snort –i etho" >> $HOME/.profile
echo "00 12 * * * $HOME/tripshell.sh > /dev/null 2>&1" >> /var/spool/cron/tabs/ninad
echo "00 12 * * * $HOME/antivirus.sh > /dev/null 2>&1" >> /var/spool/cron/tabs/ninad
echo "00 17 * * 5 $HOME/bin/backup.sh > /dev/null 2>&1" >> /var/spool/cron/tabs/ninad
echo "00 12 * * * $HOME/chkrootkit.sh > /dev/null 2>&1" >> /var/spool/cron/tabs/ninad
```

*A. Tripwire File Integrity Checking*

The system then scans the report generated by tripwire and matches the modified files to this list. If a match is found, the system requests that the administrator replace that particular file.

```
#! /bin/sh
# tripshell.sh
nice tripwire -–check
for file in `grep "Modified object name:" /tmp/report.txt | cut -c 24-`
do
grep `basename "$file"` /$HOME/severity_levels.txt > /dev/null 2>&1
if [ $? -eq 0 ]
then
echo "Critical File: $file modified!"
echo "Please replace $file with a clean backed up version"
fi
done
exit 0
```

Our system scans the file system automatically every day after noon, when the load on the machine is relatively less. This is done by adding tripshell.sh to cron jobs.

    00 12 * * * $HOME/tripshell.sh > /dev/null 2>&1

*B. Snort-IDS*

Our system blocks attacks with the help of snort. This is done by adding the rogue IP address to Iptables and specifying that it be blocked from attempting to communicate with our machine. It also effectively prevents attacks by the same attacker.

```
#! /bin/sh
grep -i -e " Unknown Traffic" -e "information leak" -e "non-standard protocol or event" -e "scan" -e "Trojan" -e "Denial of service" -e "attack" -e "suspicious" -e "system call" -e "Misc activity" -e "vulnerable" -e "privilege gain" -e "client was using an unusual port" -e "executable code was detected" -e "information leak" -e "bad traffic" -e "unusal" -e "violation" -e "Attempt to login by a default username and password" -e "Decode of an RPC Query"  /var/log/snort/alerts.ids >> /var/log/snort/scanreport

cut -d ':' -f7 /var/log/snort/scanreport|cut -c11-25|sort -u|grep -v "192.168.45.203"|grep -v "-"|sort -u -b |cat -n > blockaddress

files=$(cat blockaddress)

for var in $files
do
 echo "i am inside"
   echo "$var"
   iptables -A INPUT -s $var -p icmp --icmp-type echo-request -m length --length 1500:0xffff -j DROP
route add -host $var reject
done

echo "the following addresses have been blocked permanently"
cat blockaddress
```

*C. Chkrootkit*

chkrootkit is a tool to locally check for signs of a rootkit. It contains a chkrootkit: shell script that checks system binaries for rootkit modification. The following tests are made: aliens, dirname, echo, egrep, env, find, fingerd, gpm, grep, hdparm, su, ifconfig, inetd, inetdconf, rpcinfo, rlogind, rshd, slogin, sendmail, sshd, syslogd, tar, tcpd, top, telnetd, timed, traceroute, and write. ifpromisc.c checks whether the interface is in promiscuous mode, chklastlog.c checks for lastlog deletions, chkwtmp.c checks for wtmp deletions, check_wtmpx.c checks for wtmpx deletions (Solaris only), and chkproc.c checks for signs of LKM Trojans. The code snippet below shows that the above tests are carried out and if a rootkit has been found then the sys admin(root) is mailed an alert.

```
chk_rootkit() {

  echo checking for rootkits...
  $safe_chkrootkit | $safe_grep -e INFECTED -e Vulnerable

  if [ $? -ne 0 ]; then

    echo -e \\a ${0}: NO SECURITY ALERT: NO possible rootkit infection detected!
    # email alert message
    echo "message from ${0}: NO possible rootkit infection detected! Run ${safe_chkrootkit} to see which files are infected"\
    | $safe_mail -s "=== NO SECURITY ALERT: possible rootkit infection detected!" $mail_to

    exit 1
  fi
  exit 0
```

}

D. *Clam Antivirus*

After configuring clamscan and freshclam by doing appropriate modification to /etc/clamav/freshclam.conf, the anti-virus software is all ready to be used. All that remains to be done is to run the antivirus scan periodically, and ensure that the virus signatures database is regularly updated. Our system ensures this by scanning the system every day after noon, when the machine has less load. This is done by adding the following line to crontabs,
   00 12 * * * $HOME/antivirus.sh > /dev/null 2>&1
   The script antivirus.sh looks like this,

```
#! /bin/sh
# antivirus.sh

nice freshclam
nice clamscan --no-summary –il $HOME/antivirus_infected_list –r /usr 2> /dev/null
nice clamscan --no-summary –il $HOME/antivirus_infected_list –r /bin 2> /dev/null
nice clamscan --no-summary –il $HOME/antivirus_infected_list –r /etc 2> /dev/null
nice clamscan --no-summary –il $HOME/antivirus_infected_list –r /boot 2> /dev/null

for file in `cut –d ":" –f1 $HOME/antivirus_infected_list`
do
grep `basename "$file"` $HOME/severity_levels.txt > /dev/null 2>&1
if [ $? -eq 0 ]
then
echo "Critical File: $file infected!"
echo "Please replace $file with a clean backed up version"
else
echo "$file is infected!"
rm –i $file
fi
done
exit 0
```

*E. Evidence Gathering*

Various tools are available to gather data from a partition or a Hard drive, to simplify the process we used a built-in utility called 'dd' because it is available in all the Linux distributions.
The 'dd' command was also used because it creates an image of a partition without modifying the access times of that partition and its files, which is extremely important from the point of view of gathering evidence and not tampering with it. Below is the code snippet for creating a disk image of the partition named '/dev/sda1' to '/home/jai/data.dd'.
dd   if=/dev/sda1   of=/home/jai/data.dd   bs=4096   \ conv=notrunc,noerror
Here :

if = input file
of = output file
bs = bytes
conv = Convert the file as specified by the CONVERSION argument(s). Here notrunc means don't truncate files and noerror means continue after read errors.

*F. Evidence Analysis*

The sleuth kit include a lot of functions related to File system analysis, Data recovery, and Timeline creation & analysis. Creating a timeline of system activity will give an investigator clues regarding where to probe further. The timelines in The Sleuth Kit allow one to quickly get a high-level look at system activity, such as when files were compiled and when archives were opened. TSK allows you to generate timelines of activity from a variety of sources.We will analyze the disk image created earlier using the 'fls' and 'mactime' included in the sleuth kit.
The 'fls' command requires the '-m' argument with the '-r' flag to gather all files. This step walks through the directory hierarchy and outputs a line for each file in the file system. This command needs to be run for each partition in a disk image.
fls -m / -r home/ihrs/data.dd > bodyline.txt
#The output of the above command goes to bodyline.txt in the
#standard bodyline format of TSK 2.0
mactime -b bodyline.txt > timeline.txt
#the output of the above command goes to timeline.txt which
#contains MAC times for every file in the disk image from here #on we can use 'grep' to find out MAC times for suspected files.
#M=Modified, A=access, C=created times.
#sample grep command below
grep "\/dev\/" timeline.txt > timeline.dev.txt
#this command creates a MAC times for the /dev directory in the #partition

G. *SSH Backups*

The backup commands in our system use SSH to encrypt data over network and rsync to copy and sync files over the network.The script is added to crontab to automate the time of back ups at user discretion. SSH allows automated remote login with private key- public key encryption and is hence used in our system.

Key pair generation
On the machine you want to log in to, type ssh-keygen -t to generate the key pair.
$ ssh-keygen -t rsa
Generating public/private dsa key pair.
Enter passphrase (empty for no passphrase): [press enter here]
Enter same passphrase again: [press enter here]
Your identification has been saved in /home/bob/.ssh
Your public key has been saved in /home/bob/id_rsa.pub.
The key fingerprint is:
2e:28:d9:ec:85:21:e7:ff:73:df:2e:07:78:f0:d0:a0 bob@linux

Now copy the private key to the to the .ssh directory of the account on the machine you will be logging in from.

scp id_dsa root@foo:~/.ssh/id_dsa

Also copy the public key to the allowed list of ssh logins.

cp id_dsa.pub authorized_keys2

To enable passwordless login type

ssh-add

This sets up remote login between client machine and backup server.
Now use rsync to send back up data over the network.

rsync -avz /home/jai/bin -e ssh jai@server:/home/jai/bin
sending incremental file list
.....
watashi/

watashi/bleach_watashi_001.jpg

watashi/bleach_watashi_002.jpg

watashi/bleach_watashi_003.jpg

watashi/bleach_watashi_004.jpg

watashi/bleach_watashi_005.jpg

Here jai@server is the user on server which has read permissions to /home/jai/bin and bin dir from client side is sent to jai@server. We have automated the script by adding it as a cron job i.e. Backups occur every Friday at 5 pm.

00 17 * * 5 $HOME/bin/backup.sh > /dev/null 2>&1

## IV. TESTING

The project was deployed on a computer in a college network and tested against a variety of attacks. The software worked successfully and was able to block majority of the attacks. For example, we performed an intense port-scan attack using N-map software on a computer before and after installing our software on the machine. N-map is used by a hacker or an attacker for OS-fingerprinting in which he tries to find out the vulnerable ports that are open, the operating system of the victim's machine and other details that are required to launch an attack. After performing the test, we got the following result.

Figure 5. Nmap Output Before Automation Of System

Figure 6. Nmap Output After Automation Of System

The above two figures are the screenshots taken from the computer from which we launched the portscan attack on the victim machine using Nmap portscanner. As we can see from figure 1, Nmap is able to detect an open port which has the SSH service running on the victim computer. The attacker can use this knowledge to devise ways to penetrate into the victim's machine through the open ports. But when we install and deploy our project on the same machine, the output of Nmap indicates that all ports on the victim machine are filtered. This is because Nmap cannot determine whether the port is open because packet filtering prevents its probes from reaching the port. This forces Nmap to retry several times just in case the probe was dropped due to network congestion rather than filtering. This slows down the scan dramatically making things difficult for the attacker.

The automated system is able to block a majority of the attacks. This is because the Snort-IDS uses classification.conf file which is able to classify attacks based on their types. The various categories include denial of service, attempted/successful privilege gain, detection of network scan, Trojan activity, suspicious login, ping of death attack, flooding attack, malicious codes and other attacks.

The backups are taken care of by the SSH protocol wherein the data is transferred in the encrypted form between client and the backup server.

## V. CONCLUSION

We have thus implemented a comprehensive Incidence Handling and Security System, which is capable of handling most of the common security threats. The system covers incidents specified in the scope of the project. The system has been successfully deployed and tested on a machine running Linux Fedora 10, which uses Linux kernel version 2.6.27. The system uses the third party open source tools and software as specified before, and the programming capabilities of the Linux bash shell.

The system can be deployed on any machine working on Linux operating system. However, the component systems should have a variant for that flavor of Linux. Also, the component systems required to be pre-installed on the machine.

## VI. FUTURE RESEARCH

The automated system that has been designed is capable of handling all major basic types of attacks. But it is bounded by the limitations of the classification.conf file of the Snort-IDS. It will act on only those attacks which have been listed in the classification file. Though the list of attacks classified in the file should suffice to block a majority of attacks and keep the system secure, we need to add a capability that will make the system immune to blended attacks. Also ,with the wide variety of attacking tools being easily available today, blended attacks become easy and they pose a threat to the system. Since the performance of our automated system is decided by the components that have been integrated, care has been taken to configure all the components properly so that the overall system offers maximum protection with the best of its capabilities.

The components that have been used must be updated with time regularly for best results. There is a scope for addition of more components or replacement of any component with a new one that comes in the market with better services .Although centralized logging has been implemented to help forensics, there is a scope for the development of the customized event-correlator and system auditing as well.